\documentclass[a4paper,fleqn,useAMS,usenatbib]{mnras}

\usepackage[T1]{fontenc}
\usepackage{ae,aecompl}

\usepackage{amssymb}
\usepackage{aas_macros}
\usepackage{natbib}
\usepackage{times}
\usepackage{graphicx}
\usepackage{xspace}

\newcommand{\sub}[1]{_{\rm #1}}
\newcommand{\up}[1]{^{\rm #1}}

\newcommand{\alpharatio}{$[\alpha/\mathrm{Fe}]\sub{\ast}$\xspace}
\newcommand{\oratio}{$[\mathrm{O}/\mathrm{Fe}]\sub{\ast}$\xspace}

\defcitealias{schaye_2015}{S15}

\title[Galaxy $\alpha$-enhancement]{The origin of the $\alpha$-enhancement of massive galaxies}

\author[M. C. Segers et al.]{
Marijke C. Segers,$^{1}$\thanks{E-mail: segers@strw.leidenuniv.nl}
Joop Schaye,$^{1}$
Richard G. Bower,$^{2}$
Robert A. Crain,$^{3}$
\newauthor Matthieu Schaller$^{2}$
and Tom Theuns$^{2}$
\\
$^{1}$Leiden Observatory, Leiden University, PO Box 9513, NL-2300 RA Leiden, the Netherlands\\
$^{2}$Institute for Computational Cosmology, Department of Physics, University of Durham, South Road, Durham DH1 3LE, UK\\
$^{3}$Astrophysics Research Institute, Liverpool John Moores University, 146 Brownlow Hill, Liverpool, L3 5RF, UK
}

\date{Accepted XXX. Received YYY; in original form ZZZ}

\pubyear{2016}

\begin{document}
\label{firstpage}
\pagerange{\pageref{firstpage}--\pageref{lastpage}}
\maketitle

\begin{abstract}
We study the origin of the stellar $\alpha$-element-to-iron abundance ratio, \alpharatio, of present-day central galaxies, using cosmological, hydrodynamical simulations from the Evolution and Assembly of GaLaxies and their Environments (EAGLE) project. For galaxies with stellar masses of $M\sub{\ast} > 10^{10.5}$ M$\sub{\odot}$, \alpharatio increases with increasing galaxy stellar mass and age. These trends are in good agreement with observations of early-type galaxies, and are consistent with a `downsizing' galaxy formation scenario: more massive galaxies have formed the bulk of their stars earlier and more rapidly, hence from an interstellar medium that was mostly $\alpha$-enriched by massive stars. In the absence of feedback from active galactic nuclei (AGN), however, \alpharatio in $M\sub{\ast} > 10^{10.5}$ M$\sub{\odot}$ galaxies is roughly constant with stellar mass and decreases with mean stellar age, extending the trends found for lower-mass galaxies in both simulations with and without AGN. We conclude that AGN feedback can account for the $\alpha$-enhancement of massive galaxies, as it suppresses their star formation, quenching more massive galaxies at earlier times, thereby preventing the iron from longer-lived intermediate-mass stars (supernova Type Ia) from being incorporated into younger stars.
\end{abstract}

\begin{keywords}
galaxies: abundances -- galaxies: formation -- galaxies: star formation
\end{keywords}


\section{Introduction}
\label{sec:introduction}

The elemental abundances of galaxies hold valuable information about their formation and evolution. Galaxies build up their stellar content from gas accreted from the intergalactic medium (IGM) \citep[e.g.][]{keres_2005,vandevoort_2011} and by recycling the stellar mass released by evolved stellar populations \citep[e.g.][]{kennicutt_1994,leitner+kravtsov_2011,segers_2016}. While the former is predominantly metal-poor, stellar ejecta are intrinsically metal-rich. Stellar populations enrich the interstellar medium (ISM) by means of supernova (SN) explosions and winds from massive and asymptotic giant branch (AGB) stars, where the ejecta from each channel are characterized by a distinctive abundance pattern and are released into the ISM on different timescales. However, IGM accretion dilutes the enriched ISM, and galactic outflows can remove enriched gas from the ISM, thereby altering its chemical content. Since stars retain the abundances of the ISM in which they were formed, the stellar abundances of galaxies represent a fossil record of their formation history.

In particular, core collapse SNe release mostly $\alpha$-elements, i.e. elements that are built from the progressive addition of He ($\alpha$) particles -- including O, Ne, Mg, and Si \citep[e.g.][]{woosley+weaver_1995}. In contrast, SN Type Ia ejecta, which are released with a temporal delay relative to core collapse SN ejecta, consist almost entirely of Fe \citep[e.g.][]{thielemann_2003}. As a consequence, the O/Fe mass ratio in the ejecta of a simple stellar population generally decreases with time (see fig. 3 of \citealt{wiersma_2009b}: from $10$ Myr to $10$ Gyr, this ratio decreases by about $0.3$ dex). The $\alpha$-element-to-iron abundance ratio (also referred to as $\alpha$-enhancement, when it is compared to the solar abundance ratio) has therefore been used extensively to study the formation history of local, mainly early-type, galaxies \citep[e.g.][]{trager_2000,thomas_2010,johansson_2012,conroy_2014}. A common finding of these studies is that the $\alpha$-enhancement increases with increasing galaxy stellar mass or velocity dispersion, which is generally interpreted as evidence for galactic `downsizing'. In this galaxy formation scenario, massive galaxies form the bulk of their stars earlier and over a shorter period of time than low-mass galaxies \citep[e.g.][]{cowie_1996,neistein_2006}, which is reflected by their chemical content: due to their short formation timescales, their stars are primarily enriched in $\alpha$-elements, released by massive stars on short timescales. The crucial element in the downsizing scenario is a mechanism that efficiently quenches star formation in massive galaxies, and in such a way that more massive galaxies are quenched at earlier times. An obvious candidate is the energy feedback from active galactic nuclei \citep[AGN; e.g.][]{dimatteo_2005}.

In this Letter, we use simulations from the Evolution and Assembly of GaLaxies and their Environments (EAGLE) project (\citealt{schaye_2015}, hereafter S15; \citealt{crain_2015}) to 
study the relation of stellar $\alpha$-enhancement with galaxy stellar mass and age of central galaxies at $z = 0$, and investigate the role played by AGN feedback in reproducing the trends observed for massive, early-type galaxies. The fiducial EAGLE `reference' model reproduces key observational galaxy properties such as the evolution of the galaxy stellar mass function \citep{furlong_2015a}, the sizes of active and passive galaxies and their evolution \citep{furlong_2015b}, and the distribution of galaxies in the colour-magnitude diagram \citep{trayford_2015}. The present-day mass-metallicity relation is reproduced over the mass range of interest here ($M\sub{\ast} \gtrsim 10^{10}$ M$\sub{\odot}$; S15) (and higher-resolution EAGLE simulations of smaller volumes extend this agreement to lower masses; S15). In addition to the reference model, we use a model variation in which AGN feedback has been turned off to isolate its effect on galaxy $\alpha$-enhancement. In Section~\ref{sec:simulations} we describe the simulation set-up and implemented subgrid physics. In Section~\ref{sec:results} we present galaxy $\alpha$-enhancement as a function of stellar mass and age, and show that AGN feedback is responsible for the trends that are also observed for massive, early-type galaxies. We give our conclusions in Section~\ref{sec:conclusions}.


\section{Simulations}
\label{sec:simulations}

The EAGLE simulations were run with a modified version of the smoothed particle hydrodynamics (SPH) code \textsc{Gadget3} \citep[last described by][]{springel_2005}. Changes include the use of a pressure-entropy formulation of SPH (\citealt{hopkins_2013}; see also \citealt{schaller_2015}) and the time-step limiter of \citet{durier+dallavecchia_2012}. The simulations adopt a $\Lambda$ cold dark matter cosmology with parameters taken from \citet{planck_2014}: $\left[ \Omega\sub{m},\Omega\sub{b},\Omega\sub{\Lambda},\sigma\sub{8},n\sub{s},h \right]=\left[ 0.307,0.04825,0.693,0.8288,0.9611,0.6777 \right]$.

To investigate the effect of AGN feedback on galaxy $\alpha$-enhancement, we compare the results from two simulations: the EAGLE reference model (denoted by \emph{Ref} in S15) and a model for which the AGN feedback subgrid implementation has been turned off, while keeping all the other subgrid parameters the same (\emph{NoAGN}). These were run in periodic volumes of size $L = 50$ comoving Mpc, containing $N = 752\up{3}$ dark matter particles with mass $m\sub{dm}=9.7 \times 10^6\ {\rm M}\sub{\odot}$ and an equal number of baryonic particles with initial mass $m\sub{b}=1.8 \times 10^6\ {\rm M}\sub{\odot}$. The gravitational softening length is set to $2.66$ comoving kpc and limited to a maximum of $0.7$ proper kpc at low redshift. In addition, since we are primarily concerned with high-mass ($M\sub{\ast} > 10^{10}$ M$\sub{\odot}$) galaxies, we also use the $L = 100$ Mpc, $N = 1504\up{3}$ (hence identical mass resolution and softening length) simulation of the reference model to improve the sampling of the massive galaxy population.

The simulations include a number of subgrid models for physical processes that originate on unresolved scales. These include star formation, which is modelled with a metallicity-dependent threshold \citep{schaye_2004} and a pressure-dependent star formation law that reproduces the Kennicutt-Schmidt relation \citep{schaye+dallavecchia_2008}. Star particles represent stellar populations of a single age, with their mass (distributed according to the \citealt{chabrier_2003} initial mass function; IMF) and metallicity inherited from their progenitor gas particles. The simulations follow the abundances of $11$ elements (including O and Fe) as they are gradually released into the ISM according to the prescriptions of \citet{wiersma_2009b}. These abundances are used to calculate the rates of radiative cooling and heating \citep{wiersma_2009a}. For each star particle, the rate of SN Type Ia per unit stellar mass is $\nu \mathrm{e}^{-t/\tau}$, where $t$ is the stellar age, and the parameters $\tau = 2$ Gyr and $\nu = 2 \times 10^{-3}$ M$\sub{\odot}^{-1}$ were chosen to reproduce the evolution of the observed SN Type Ia rate density (S15).

The reference model includes a prescription for the growth of black holes (BHs), which increase their mass via mergers and gas accretion, where the accretion rate depends on the angular momentum of the gas \citep[][S15]{rosasguevara_2015}. Energy feedback from star formation and AGN (the latter is omitted in the \emph{NoAGN} model) is implemented by stochastically heating gas particles surrounding newly formed star particles and BH particles, respectively \citep{dallavecchia+schaye_2012}, so that galactic winds develop naturally without turning off the radiative cooling or hydrodynamics. The subgrid parameters governing the efficiencies of stellar and AGN feedback have been calibrated to reproduce the observed present-day galaxy stellar mass function and the relation between stellar mass and BH mass, with the additional constraint that the sizes of galaxies must be reasonable \citep[S15;][]{crain_2015}.

Haloes are identified using the Friends-of-Friends (FoF) and \textsc{subfind} algorithms \citep{dolag_2009}. In this work, we are only concerned with `central' galaxies, which are the galaxies residing at the minimum of the halo potentials. Following S15, we use a spherical aperture of radius $30$ kpc to calculate galaxy properties.

We adopt the usual definition of the stellar abundance ratio,
\begin{equation}\label{eq:ab_ratio}
\left[\frac{\mathrm{O}}{\mathrm{Fe}}\right]\sub{\ast} = \log\sub{10} \left(\frac{X\up{O}}{X\up{Fe}}\right) - \log\sub{10} \left(\frac{X\sub{\odot}\up{O}}{X\sub{\odot}\up{Fe}}\right),
\end{equation}
where in our case $X^x = \Sigma\sub{i} m\sub{i}\up{O} / \Sigma\sub{i} m\sub{i}$ is the galaxy stellar mass fraction in element $x$, with $m\sub{i}\up{O}$ and $m\sub{i}$ being the oxygen and total particle masses, respectively. $X\sub{\odot}\up{O}/X\sub{\odot}\up{Fe} = 4.44$ is the solar abundance ratio \citep{asplund_2009}. Throughout this work, we will use \oratio as a proxy for \alpharatio in EAGLE, as oxygen dominates the mass fraction of $\alpha$-elements.


\section{Results}
\label{sec:results}

To investigate the effect of AGN feedback on the stellar $\alpha$-enhancement of galaxies, we first show the impact on the trend with stellar mass. Then, since $\alpha$-enhancement is often used as a proxy for galaxy age (see e.g. \citealt{renzini_2006} for a review), we explore the impact of AGN feedback on the relation between \alpharatio and age.

\subsection{Relation with stellar mass}
\label{sec:results_vs_mass}

\begin{figure}
\begin{center}
\includegraphics[width=\columnwidth]{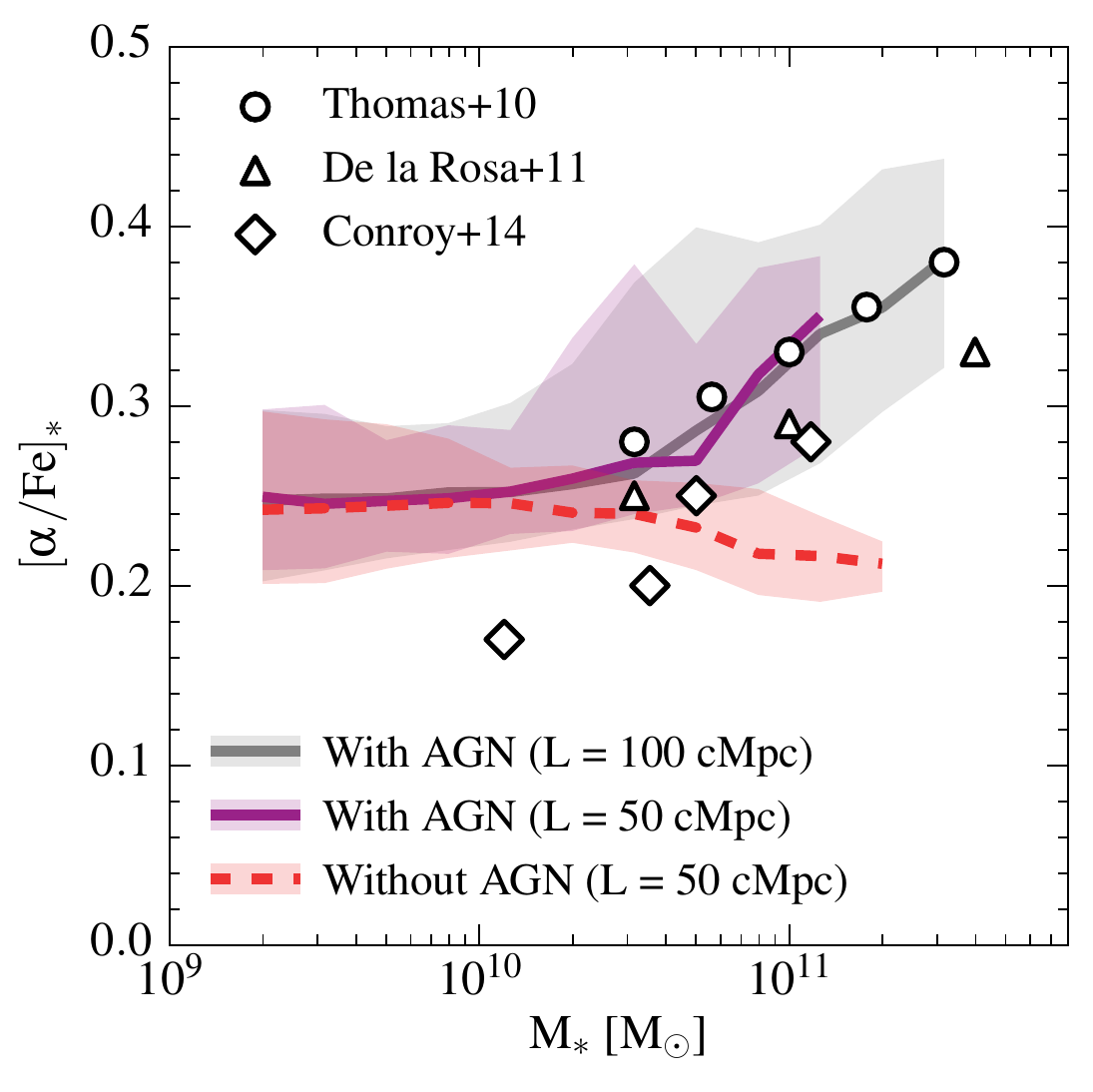}
\end{center}
\caption{The effect of AGN feedback on the $\alpha$-element-to-iron abundance ratio as a function of galaxy stellar mass for present-day central galaxies. The curves show the median \alpharatio (represented by \oratio) in logarithmic mass bins of size $0.2$ dex, as predicted by the simulations with and without AGN feedback. The shaded regions mark the $10$th to $90$th percentile range. We only show bins containing at least $10$ galaxies and corresponding to a stellar mass of at least $1000$ particles. In the fiducial model, \alpharatio increases with mass for $M\sub{\ast} \gtrsim 10^{10.5}$ M$\sub{\odot}$, in excellent agreement with the observations from \citet{thomas_2010}, and in qualitative agreement with those of \citet{delarosa_2011} and \citet{conroy_2014} (all converted to a solar abundance ratio of $X\sub{\odot}\up{O}/X\sub{\odot}\up{Fe} = 4.44$). In the model without AGN feedback, however, \alpharatio is roughly constant over the whole mass range.}
\label{fig:abratio_vs_mass}
\end{figure}

Fig.~\ref{fig:abratio_vs_mass} shows the stellar $\alpha$-enhancement, \alpharatio (represented by \oratio), as a function of galaxy stellar mass, comparing EAGLE simulations with and without AGN feedback. As we showed previously \citep{segers_2016}, in the EAGLE reference simulation \alpharatio is $\sim 0.25$ for $M\sub{\ast} \lesssim 10^{10.5}$ M$\sub{\odot}$ and increases with stellar mass for $M\sub{\ast} \gtrsim 10^{10.5}$ M$\sub{\odot}$, in excellent agreement with the observed \alpharatio - $M\sub{\ast}$ relation for early-type galaxies reported by \citet{thomas_2010}. This trend is also in qualitative agreement with the observations from \citet{delarosa_2011} and \citet{conroy_2014}. For the latter, we only include their measurements at $M\sub{\ast} > 10^{10}$ M$\sub{\odot}$, where EAGLE galaxies are predominantly early type \citep[S15;][]{trayford_2015}. The relation between \alpharatio and stellar velocity dispersion, $\sigma\sub{\ast}$, matches the observations equally well. We show the relation with stellar mass here, since the simulation includes late-type galaxies for which rotation may influence the velocity dispersion. In addition, we explore the effect of calculating \alpharatio as a luminosity-weighted average, using the r-band luminosities of EAGLE star particles as computed by \citet{trayford_2015}, and in the absence of dust, instead of a ratio of mass fractions (equation~\ref{eq:ab_ratio}). We find that the difference is marginal, typically within $\pm 0.05$ dex, and that it leaves the high-mass end slope unchanged.

We note that the normalization of the simulated relation is uncertain by a factor of $\sim 2$ due to uncertainties in the nucleosynthetic yields and SN Type Ia rate \citep{wiersma_2009b}, while the normalization of the observed relations varies as a result of uncertainties in the stellar population modelling. The systematic uncertainty in the observed \alpharatio is typically estimated to be $\sim 0.05 - 0.1$ dex \citep{thomas_2010,delarosa_2011,johansson_2012,conroy_2014}.

The reference model ran in the smaller volume ($L = 50$ Mpc) yields results consistent with the $L = 100$ Mpc simulation. Comparing this model to the one without AGN feedback, we find that in the absence of AGN feedback \alpharatio does not increase with stellar mass. The relation in fact exhibits a mildly negative gradient. From this we infer that feedback from AGN, becoming effective at $M\sub{\ast} \sim 10^{10.5}$ M$\sub{\odot}$, is responsible for the $\alpha$-enhancement of present-day massive galaxies. It suppresses star formation in the progenitor galaxies, when these progenitors and their central BHs have grown massive enough for AGN feedback to be efficient, which happens earlier for more massive galaxies. Hence, since most stars formed before AGN feedback became active, these galaxies have earlier and shorter formation times, which naturally leads to enhanced enrichment of the stellar phase by ejecta from short-lived massive stars. In the absence of this quenching mechanism, late star formation is not suppressed efficiently enough to prevent the incorporation of iron from longer-lived intermediate-mass stars (mainly SN Type Ia) into secondary generations of stars. Considering the timescale on which the Fe from SN Type Ia is released, we infer that AGN must significantly suppress star formation within $\lesssim 1$ Gyr.

Reproducing the observed relation between \alpharatio and stellar mass (or, equivalently, stellar velocity dispersion) for early-type galaxies has been a problem for many models of galaxy formation \citep[e.g.][]{nagashima_2005,pipino_2009}, although recently a few semi-analytic models have been more successful \citep{arrigoni_2010,yates_2013}. While it has been argued that a variable IMF is necessary to reproduce the observed trend \citep{calura+menci_2009,gargiulo_2015}, our results and those of \citet{yates_2013} show that this can also be achieved with a universal Chabrier IMF. Our model comparison shows that the inclusion of AGN feedback is sufficient to generate a sufficiently steep \alpharatio - $M\sub{\ast}$ relation. The effect of AGN feedback on this relation has been demonstrated before, by \citet{calura+menci_2011} using a semi-analytic model that broadly reproduces the observed \alpharatio - $\sigma$ relation, if both AGN feedback and interaction triggered starbursts are included. Furthermore, using low-resolution hydrodynamical simulations, \citet{taylor_2015} find an effect on two example galaxies that is in qualitative agreement with our results. However, in contrast to this work, the effect on the overall (massive) galaxy population is small, and their implementation of AGN feedback does not reproduce the observed trend.

\subsection{Relation with stellar age}
\label{sec:results_vs_age}

\begin{figure}
\begin{center}
\includegraphics[width=\columnwidth]{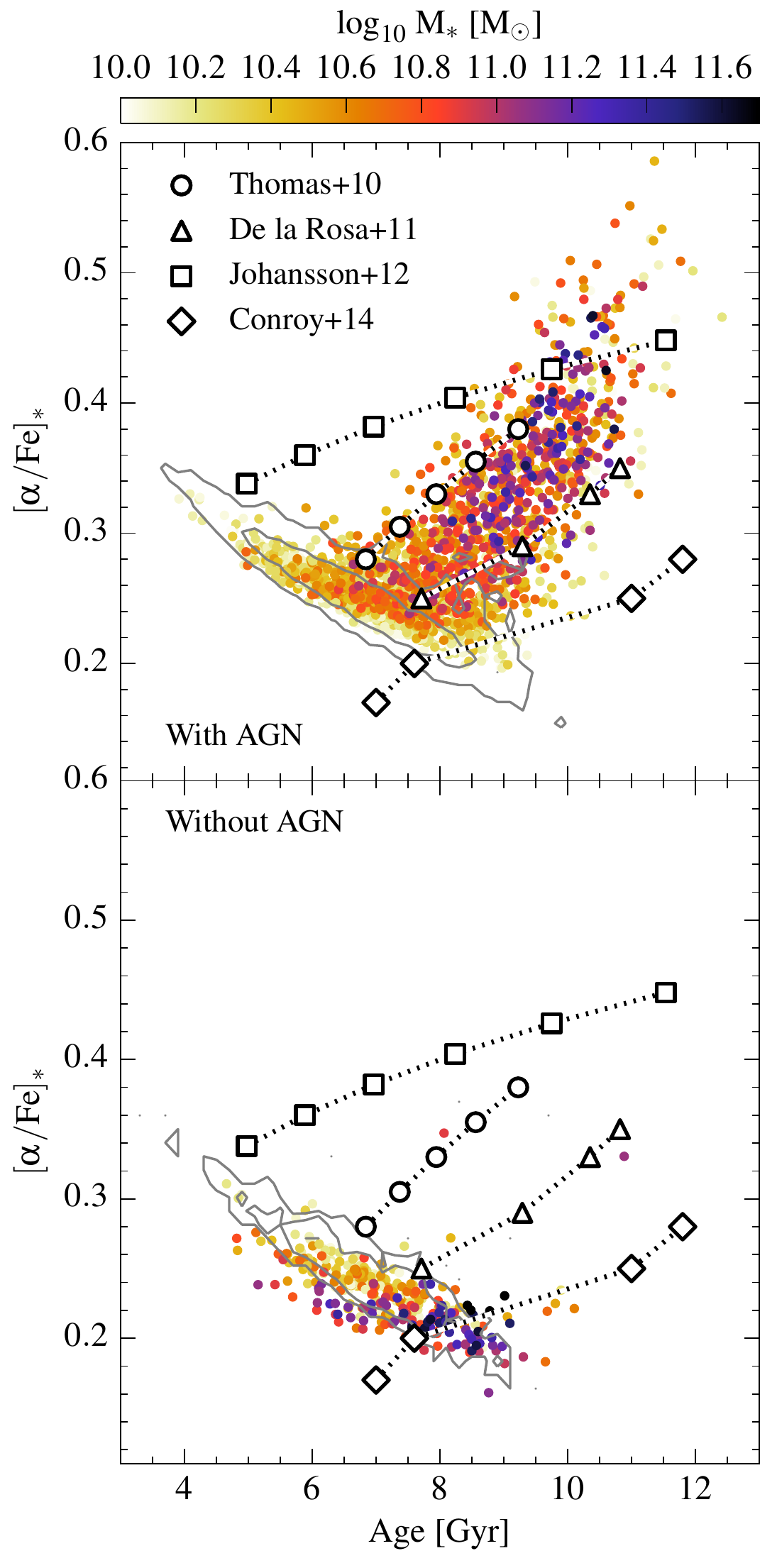}
\end{center}
\caption{The effect of AGN feedback on the relation between \alpharatio and mass-weighted mean stellar age. The top and bottom panels show the results from the simulations with ($L = 100$ Mpc) and without ($L = 50$ Mpc) AGN feedback, respectively. Galaxies with $M\sub{\ast} \geq 10^{10}$ M$\sub{\odot}$ are shown as filled circles, coloured by their stellar mass, while the distribution for $1.8 \times 10^9$~M$\sub{\odot} < M\sub{\ast} < 10^{10}$~M$\sub{\odot}$ galaxies is shown as grey contours enclosing the $68$th and $95$th percentiles. There are fewer objects in the bottom panel due to the smaller simulation volume. In both models, galaxies with $M\sub{\ast} \lesssim 10^{10.5}$ M$\sub{\odot}$ have lower \alpharatio with increasing age, since they are increasingly enriching their ISM and stars with Fe as they continue to form stars. In contrast, because star formation in galaxies with $M\sub{\ast} \gtrsim 10^{10.5}$ M$\sub{\odot}$ is quenched, they are less Fe-enriched when they are older, in qualitative agreement with observations of early-type galaxies \citep[][all converted to a solar abundance ratio of $X\sub{\odot}\up{O}/X\sub{\odot}\up{Fe} = 4.44$]{thomas_2010,delarosa_2011,johansson_2012,conroy_2014}. In the absence of AGN feedback, however, the $M\sub{\ast} \gtrsim 10^{10.5}$ M$\sub{\odot}$ galaxies follow the trend of the lower-mass galaxies, decreasing their \alpharatio with time as they continue to form stars.}
\label{fig:abratio_vs_age}
\end{figure}

\begin{figure}
\begin{center}
\includegraphics[width=\columnwidth]{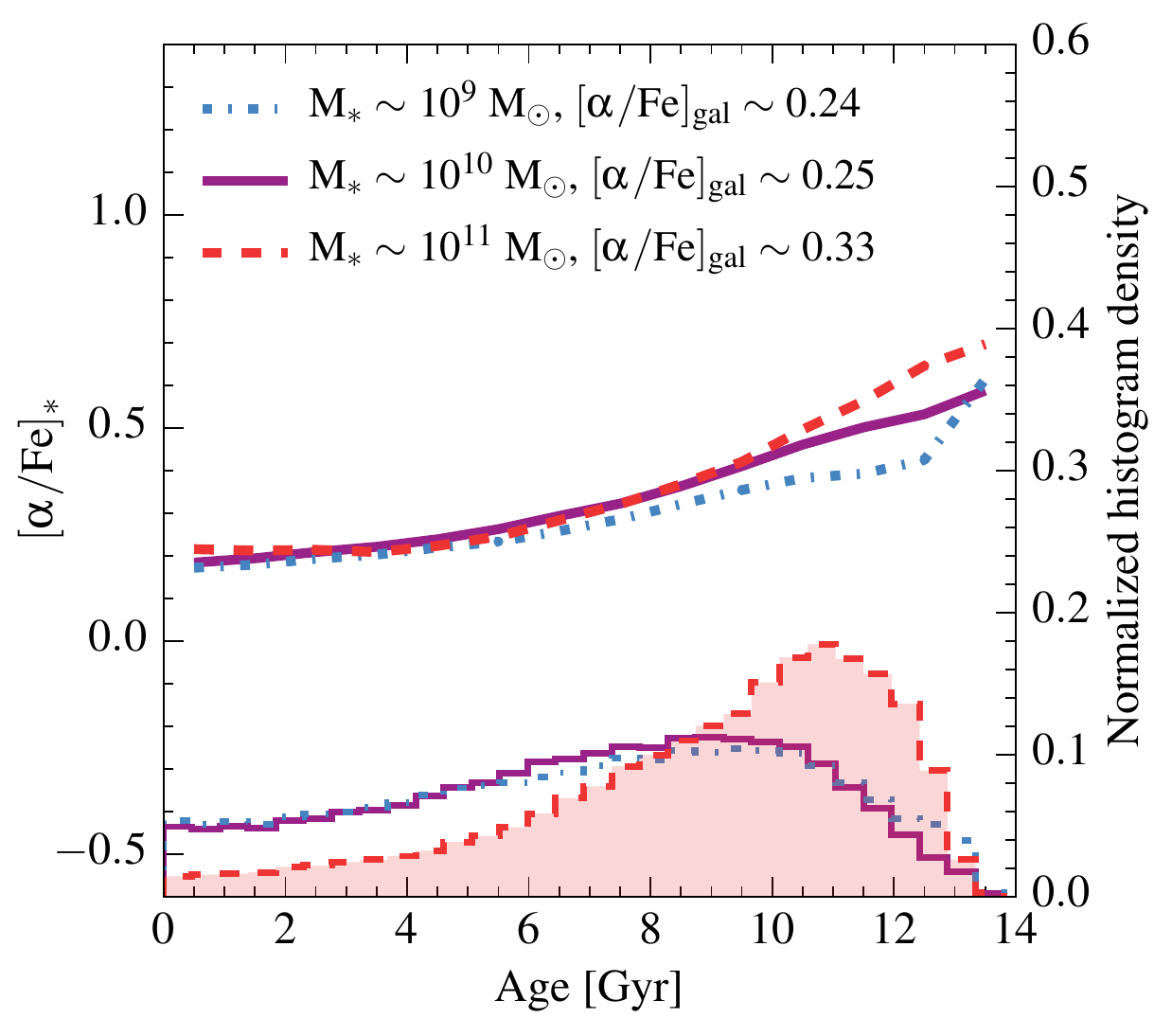}
\end{center}
\caption{The relation between \alpharatio and age for individual star particles in galaxies with stellar masses of $M\sub{\ast} / \mathrm{M}\sub{\odot} \sim 10^{9}, 10^{10}, 10^{11}$, selected from the $L = 100$ Mpc reference simulation. The $0.1$ dex galaxy mass bins contain $748$, $300$, $92$ galaxies, respectively. The upper curves show, for all the star particles in the galaxies of the respective mass bin, the median \alpharatio in each $\Delta$age = $1$ Gyr bin with at least $10$ particles, while the (normalized) histograms at the bottom show the distributions of star particle ages, weighted by their current mass. While the relation between \alpharatio and age of individual star particles is similar for all three galaxy mass bins, the different age distributions lead to different values of the galactic \alpharatio.}
\label{fig:abratio_vs_age_stars}
\end{figure}

If AGN feedback is the mechanism that quenches star formation in galaxies, turning off AGN feedback should affect the ages of galaxies. Before comparing models with and without AGN feedback, we first show in the top panel of Fig.~ \ref{fig:abratio_vs_age} the relation between \alpharatio and age for the $L = 100$ Mpc reference simulation. Here, `age' refers to the average stellar age, weighted by the initial mass of the stellar population (star particle), i.e. the mass at the time it was formed. Galaxies with $M\sub{\ast} \geq 10^{10}$ M$\sub{\odot}$ are shown as filled circles, where the colour indicates their stellar mass. Galaxies with $1.8 \times 10^9$ M$\sub{\odot} < M\sub{\ast} < 10^{10}$ M$\sub{\odot}$ are shown as grey contours depicting the $68$th and $95$th percentiles of the distribution.

The two mass regimes show opposite trends. For $M\sub{\ast} < 10^{10}$~M$\sub{\odot}$, galaxies of increasing mean stellar age have lower \alpharatio. These galaxies are still forming stars from gas that becomes increasingly Fe-enriched by SN Type Ia as time progresses. Therefore, the oldest galaxies, which have had the longest time to enrich their ISM and stars with Fe, have the lowest \alpharatio. Note that the negative trend between \alpharatio and mean stellar age for low-mass galaxies, as well as the positive trend for high-mass galaxies, is consistent with a positive correlation between age and \alpharatio for \emph{individual} stars within a galaxy. This is shown in Fig.~\ref{fig:abratio_vs_age_stars} for galaxies with $M\sub{\ast} / \mathrm{M}\sub{\odot} \sim 10^{9}, 10^{10}, 10^{11}$ from the $L = 100$ Mpc reference simulation. Star particles that were formed earlier, have higher \alpharatio, since the ISM at their formation time was less enriched by SN Type Ia than at later times. The trends in Fig.~\ref{fig:abratio_vs_age_stars}, which are similar for all three galaxy mass bins, agree qualitatively with observations of stellar populations in the Milky Way \citep[e.g.][]{haywood_2013,ramirez_2013}. It is the distribution of the star particle ages, which are shown as histograms at the bottom of Fig.~\ref{fig:abratio_vs_age_stars}, which causes the median \emph{galactic} \alpharatio of $M\sub{\ast} \sim 10^{11}$ M$\sub{\odot}$ galaxies to be enhanced with respect to those of the less massive galaxies.

Galaxies with $M\sub{\ast} \geq 10^{10}$ M$\sub{\odot}$ (coloured circles in Fig.\ref{fig:abratio_vs_age}) show a transition from \alpharatio \emph{decreasing} with age ($M\sub{\ast} \lesssim 10^{10.5}$ M$\sub{\odot}$) to \alpharatio \emph{increasing} with age ($M\sub{\ast} \gtrsim 10^{10.5}$ M$\sub{\odot}$). The latter trend is consistent with the downsizing scenario for early-type galaxies, in which older galaxies have formed the bulk of their stars from a less Fe-enriched ISM, before their star formation was quenched (as illustrated by the $M\sub{\ast} \sim 10^{11}$ M$\sub{\odot}$ galaxies in Fig.~\ref{fig:abratio_vs_age_stars}). The colour coding in Fig.~\ref{fig:abratio_vs_age} indicates that more massive galaxies typically have higher ages and higher \alpharatio ratios, consistent with Fig.\ref{fig:abratio_vs_mass}.

For early-type galaxies there are a number of observational studies on ages and abundance ratios to compare with. The black symbols in Fig.~\ref{fig:abratio_vs_age} show the observed \alpharatio (or \oratio) as a function of luminosity-weighted stellar age from \citet{thomas_2010} and \citet{conroy_2014}, and \alpharatio as a function of mass-weighted stellar age from \citet{delarosa_2011}. These correspond to the same mass bins as in Fig.~\ref{fig:abratio_vs_mass}. We also show the observed relation between \oratio and luminosity-weighted age from \citet{johansson_2012}, by evaluating their best-fit relations at $6$ stellar velocity dispersions between $10\up{1.9}$ km/s and $10\up{2.4}$ km/s (spaced by $0.1$ dex). Note that the systematic offsets between the different observed relations is large. In addition to uncertainties in the stellar population modelling (systematic uncertainties in the stellar age are typically estimated to be $\sim 0.1 - 0.2$ dex), this may be due to the size of the aperture used, and the fact that luminosity-weighted ages are typically lower than mass-weighted ages, as young stars generally dominate the population luminosity \citep[e.g.][]{trager+somerville_2009}. Comparing the observed relations to the relation predicted by EAGLE, we see that they all agree qualitatively on the positive correlation between \alpharatio and age, as shown by the $M\sub{\ast} \gtrsim 10^{10.5}$ M$\sub{\odot}$ EAGLE galaxies (which are predominantly early-type galaxies). Weighting the average stellar ages in EAGLE by the r-band luminosity of the star particles (instead of their mass) yields ages that are lower by $\sim 1-2$ Gyr. However, this does not have a significant impact on the slopes of the two galaxy populations.

To illustrate the effect of AGN feedback on the ages of galaxies, we show in the bottom panel of Fig.~\ref{fig:abratio_vs_age} the relation between \alpharatio and age for the $L = 50$ Mpc simulation without AGN feedback. While the two models agree for $M\sub{\ast} \lesssim 10^{10.5}$ M$\sub{\odot}$, the increase in \alpharatio with age for $M\sub{\ast} \gtrsim 10^{10.5}$ M$\sub{\odot}$ galaxies in the reference model is absent in the model without AGN feedback. Instead, the high-mass galaxies also show an inverted \alpharatio - age relation, with a slope and normalization similar to that of the low-mass population.


\section{Conclusions}
\label{sec:conclusions}

We have investigated the effect of AGN feedback on the stellar $\alpha$-enhancement of present-day central galaxies, using cosmological simulations from the EAGLE project. We compared results from the EAGLE fiducial model, which includes energy feedback from AGN, to results from a model without AGN feedback. We found that in the presence of AGN feedback, the \alpharatio of $M\sub{\ast} \gtrsim 10^{10.5}$ M$\sub{\odot}$ galaxies increases with both increasing galaxy stellar mass and age. These trends are in good agreement with observations of early-type galaxies, and are consistent with galactic downsizing: the earlier and more rapid formation timescales of massive galaxies result in higher \alpharatio, as the bulk of their stars was formed from a mostly $\alpha$-enriched ISM, before their star formation was quenched significantly. In the model without AGN feedback, however, \alpharatio of galaxies is insensitive to stellar mass and decreases with increasing mean stellar age, following the relation found for $M\sub{\ast} \lesssim 10^{10.5}$ M$\sub{\odot}$ galaxies, also by the model that includes AGN feedback. In the absence of a quenching mechanism, galaxies continue to form stars from an increasingly Fe-enriched ISM, decreasing their \alpharatio as they age. Consistent with earlier suggestions by \citet{calura+menci_2011} and \citet{taylor_2015}, we conclude -- for the first time using cosmological, hydrodynamical simulations that successfully reproduce the relations observed for early-type galaxies -- that star formation quenching by AGN feedback can account for the $\alpha$-enhancement of massive galaxies.


\section*{Acknowledgements}

This work used the DiRAC Data Centric system at Durham University, operated by the Institute for Computational Cosmology on behalf of the STFC DiRAC HPC Facility (www.dirac.ac.uk). This equipment was funded by BIS National E-infrastructure capital grant ST/K00042X/1, STFC capital grant ST/H008519/1, and STFC DiRAC Operations grant ST/K003267/1 and Durham University. DiRAC is part of the National E-Infrastructure. We also acknowledge PRACE for access to the resource Curie at Tr\'{e}s Grand Centre de Calcul. This work received financial support from the European Research Council under the European Union's Seventh Framework Programme (FP7/2007-2013) / ERC Grant agreement 278594-GasAroundGalaxies, from the UK STFC (grant numbers ST/F001166/1 and ST/I000976/1), and from the Belgian Science Policy Office ([AP P7/08 CHARM]). RAC is a Royal Society University Research Fellow.


\bibliographystyle{mnras}
\bibliography{bibliography}

\bsp
\label{lastpage}
\end{document}